\documentstyle[wstwocl,epsfig]{article}
\begin{document}

\bibliographystyle{unsrt}    

\newcommand{\st}{\scriptstyle}
\newcommand{\sst}{\scriptscriptstyle}
\newcommand{\mco}{\multicolumn}
\newcommand{\epp}{\epsilon^{\prime}}
\newcommand{\vep}{\varepsilon}
\newcommand{\ra}{\rightarrow}
\newcommand{\ppg}{\pi^+\pi^-\gamma}
\newcommand{\vp}{{\bf p}}
\newcommand{\ko}{K^0}
\newcommand{\kb}{\bar{K^0}}
\newcommand{\al}{\alpha}
\newcommand{\ab}{\bar{\alpha}}
\def\be{\begin{equation}}
\def\ee{\end{equation}}
\def\bea{\begin{eqnarray}}
\def\eea{\end{eqnarray}}
\def\CPbar{\hbox{{\rm CP}\hskip-1.80em{/}}}

\def\ap#1#2#3   {{\em Ann. Phys. (NY)} {\bf#1} (#2) #3}
\def\apj#1#2#3  {{\em Astrophys. J.} {\bf#1} (#2) #3}
\def\apjl#1#2#3 {{\em Astrophys. J. Lett.} {\bf#1} (#2) #3}
\def\app#1#2#3  {{\em Acta. Phys. Pol.} {\bf#1} (#2) #3}
\def\ar#1#2#3   {{\em Ann. Rev. Nucl. Part. Sci.} {\bf#1} (#2) #3}
\def\cpc#1#2#3  {{\em Computer Phys. Comm.} {\bf#1} (#2) #3}
\def\err#1#2#3  {{\it Erratum} {\bf#1} (#2) #3}
\def\ib#1#2#3   {{\it ibid.} {\bf#1} (#2) #3}
\def\jmp#1#2#3  {{\em J. Math. Phys.} {\bf#1} (#2) #3}
\def\ijmp#1#2#3 {{\em Int. J. Mod. Phys.} {\bf#1} (#2) #3}
\def\jetp#1#2#3 {{\em JETP Lett.} {\bf#1} (#2) #3}
\def\jpg#1#2#3  {{\em J. Phys. G.} {\bf#1} (#2) #3}
\def\mpl#1#2#3  {{\em Mod. Phys. Lett.} {\bf#1} (#2) #3}
\def\nat#1#2#3  {{\em Nature (London)} {\bf#1} (#2) #3}
\def\nc#1#2#3   {{\em Nuovo Cim.} {\bf#1} (#2) #3}
\def\nim#1#2#3  {{\em Nucl. Instr. Meth.} {\bf#1} (#2) #3}
\def\np#1#2#3   {{\em Nucl. Phys.} {\bf#1} (#2) #3}
\def\pcps#1#2#3 {{\em Proc. Cam. Phil. Soc.} {\bf#1} (#2) #3}
\def\pl#1#2#3   {{\em Phys. Lett.} {\bf#1} (#2) #3}
\def\prep#1#2#3 {{\em Phys. Rep.} {\bf#1} (#2) #3}
\def\prev#1#2#3 {{\em Phys. Rev.} {\bf#1} (#2) #3}
\def\prl#1#2#3  {{\em Phys. Rev. Lett.} {\bf#1} (#2) #3}
\def\prs#1#2#3  {{\em Proc. Roy. Soc.} {\bf#1} (#2) #3}
\def\ptp#1#2#3  {{\em Prog. Th. Phys.} {\bf#1} (#2) #3}
\def\ps#1#2#3   {{\em Physica Scripta} {\bf#1} (#2) #3}
\def\rmp#1#2#3  {{\em Rev. Mod. Phys.} {\bf#1} (#2) #3}
\def\rpp#1#2#3  {{\em Rep. Prog. Phys.} {\bf#1} (#2) #3}
\def\sjnp#1#2#3 {{\em Sov. J. Nucl. Phys.} {\bf#1} (#2) #3}
\def\spj#1#2#3  {{\em Sov. Phys. JEPT} {\bf#1} (#2) #3}
\def\spu#1#2#3  {{\em Sov. Phys.-Usp.} {\bf#1} (#2) #3}
\def\zp#1#2#3   {{\em Zeit. Phys.} {\bf#1} (#2) #3}

\setcounter{secnumdepth}{2} 

\begin{titlepage}
\hbox to \hsize{
\hfill\vtop{\hbox{}
\vspace{5.mm}
\hbox{\large hep-ph/9510265}
\vspace{2.mm}
\hbox{\large UB-HET-95-04}
\vspace{2.mm}
\hbox{\large October 1995} } }
\vspace{30mm}
\begin{center}
{\Large \sc Self-Couplings of Electroweak Bosons: \\[2.mm]
Theoretical Aspects and
Tests at Hadron Colliders\footnote{\normalsize Talk given at the
International Europhysics Conference on High Energy Physics, Brussels,
Belgium, July~27 --~August~2, 1995}\\[1.cm]}

{\large \sc U.~Baur\\[7.mm]}

{\large \it Physics Department, State University of New York at
Buffalo\\[1.mm]
Buffalo, NY 14260, USA\\[3.cm]}
{\large \bf
ABSTRACT \\[6.mm]}
\begin{minipage}{5.5in}
{ \baselineskip=16pt \large
The current theoretical understanding of anomalous gauge boson couplings
is reviewed, and the direct measurement of these couplings
in present and future hadron collider experiments is briefly discussed.
}
\end{minipage}
\end{center}
\end{titlepage}
\newpage

\title{Self-Couplings of Electroweak Bosons: Theoretical Aspects and
Tests at Hadron Colliders}

\firstauthors{U.~Baur }

\firstaddress{Physics Department, State University of New York at
Buffalo, Buffalo, NY 14260, USA}

\twocolumn[\faketitle\abstracts{
The current theoretical understanding of anomalous gauge boson couplings
is reviewed, and the direct measurement of these couplings
in present and future hadron collider experiments is briefly discussed.
}]

\section{Introduction}
Although the electroweak Standard Model (SM) based on an
${\rm SU_L(2)}\bigotimes {\rm U_Y(1)}$
gauge theory  has been very successful
in describing contemporary high  energy physics experiments, the three
vector-boson couplings predicted by this non-Abelian gauge theory
remain largely untested experimentally.
A direct measurement of these vector boson couplings is possible in
present and future collider experiments, in particular via pair
production processes like
$e^+e^- \to W^+W^-$, $Z\gamma$ and  $q \bar q \to W^+W^-,\; W\gamma,\;
Z\gamma,\; WZ$. Here we present a short overview of the theoretical aspects
of vector boson self-couplings, and discuss their measurement in hadron
collider experiments. Tests at LEP~II and Linear Colliders are discussed
in Refs.~\ref{Sek} and~\ref{Bark}.

\section{Theoretical Aspects}
Analogous to the introduction of arbitrary vector and axial vector
couplings $g_V$ and $g_A$ for the coupling of gauge bosons to
fermions, the measurement of the $WWV$ ($V=W,\, Z$) couplings can be
made quantitative by
introducing a more general $WWV$ vertex. For our discussion of experimental
sensitivities in Section~3 we shall use a parameterization in terms of the
phenomenological effective Lagrangian\cite{HPZH} (other, equivalent,
parameterizations are possible\cite{Bil}):
\begin{eqnarray}
\noalign{\vskip -5pt}
i{\cal L}_{WWV} & \!\!\!= & \!\!\!g_{WWV}\,
\Biggl[ g_1^V \bigl( W_{\mu\nu}^{\dagger} W^{\mu} V^{\nu}
                  -W_{\mu}^{\dagger} V_{\nu} W^{\mu\nu} \bigr) \nonumber
                  \\
& & \!\!\!+ \kappa_V W_{\mu}^{\dagger} W_{\nu} V^{\mu\nu}
+ {\lambda_V \over m_W^2} W_{\lambda \mu}^{\dagger} W^{\mu}_{\nu}
V^{\nu\lambda}
\Biggr] \>.
\label{EQ:LAGRANGE}
\end{eqnarray}
Here the overall coupling constants are defined as $g_{WW\gamma}=e$ and
$g_{WWZ}= e \cot\theta_W$, $W_{\mu\nu}= \partial_\mu W_\nu- \partial_\nu
W_\mu$, and $V_{\mu\nu}=\partial_\mu V_\nu - \partial_\nu V_\mu$.
Within the SM, at tree level, the couplings are given by
$g_1^Z = g_1^\gamma = \kappa_Z = \kappa_\gamma = 1,\; \lambda_Z =
\lambda_\gamma = 0$. $g_1^\gamma=1$ is fixed by electromagnetic gauge
invariance. $g_1^Z$,
however, may well be different from its SM value 1 and appears at the
same level as $\kappa_\gamma$ or $\kappa_Z$.

The effective Lagrangian of Eq.~(\ref{EQ:LAGRANGE}) parameterizes the
most general
Lorentz invariant and $C$ and $P$ conserving $WWV$ vertex which can be
observed in
processes where the vector bosons couple to effectively massless
fermions. Terms
with higher derivatives are equivalent to a dependence of the couplings on the
vector boson momenta and thus merely lead to a form-factor behaviour of these
couplings. Analogous to the general $WWV$ vertex it is possible to
parameterize anomalous $Z\gamma V,\, V=\gamma ,Z$ couplings in terms of
two coupling constants, $h_3^V$ and $h_4^V$ if $CP$ invariance is
imposed. All $Z\gamma V$ couplings are $C$ odd.

In the absence of a specific model of new physics, effective Lagrangian
techniques are extremely useful. An effective Lagrangian
parameterizes, in a model-independent way, the low-energy effects of the new
physics to be found at higher energies. It is only necessary to specify the
particle content and the symmetries of the low-energy theory. Although
effective Lagrangians contain an infinite number of terms, they are
organized in powers of $1/\Lambda$, where $\Lambda$ is the scale of new
physics. Thus, at energies which are much smaller than $\Lambda$, only the
first few terms of the effective Lagrangian are important. Since  all
experimental evidence is consistent with the existence of an
${\rm SU_L(2)}\bigotimes {\rm U_Y(1)}$
gauge symmetry it is natural to require the effective Lagrangian
describing anomalous gauge boson couplings to possess this invariance.
How this symmetry
is  realized depends on the particle content of the effective Lagrangian.
If one  includes a Higgs boson, the symmetry can be realized linearly,
otherwise a  nonlinear realization of the gauge symmetry is required.

If the ${\rm SU_L(2)}\bigotimes {\rm U_Y(1)}$ symmetry is realized
linearly and the analysis is restricted to operators of dimension~6,
three operators give rise to anomalous triple gauge boson
couplings\cite{Buch}:
\begin{equation}
{\cal L}_{eff}={1\over\Lambda^2}\left[f_B{\cal O}_B+f_W{\cal O}_W+
f_{WWW}{\cal O}_{WWW}\right] .
\label{ops}
\end{equation}
The explicit form of the three operators ${\cal O}_B$, ${\cal O}_W$, and
${\cal O}_{WWW}$ can be found in Ref.~\ref{HISZ}. If
$f_B=f_W$, the number of independent $WWV$ couplings is reduced to two.
Choosing $\kappa_\gamma$ and $\lambda_\gamma$ as independent parameters,
the $WWZ$ couplings are then given by (``HISZ scenario''\cite{HISZ}):
\begin{eqnarray}
\Delta g_1^Z & = & {1\over 2\cos^2\theta_W}\,\Delta\kappa_\gamma ,
\label{EQ:HISZ1} \\[1.mm]
\Delta\kappa_Z & = & {1\over 2}\,(1-\tan^2\theta_W)\,
\Delta\kappa_\gamma , \label{EQ:HISZ2}
\\[1.mm]
\lambda_Z & = & \lambda_\gamma ,
\label{EQ:HISZ3}
\end{eqnarray}
where $\theta_W$ is the weak mixing angle, $\Delta\kappa_V=\kappa_V-1$
and $\Delta g_1^Z = g_1^Z -1$.
However, it should be noted that these relations are modified when
operators of dimension~8 or higher are included\cite{HISZ}.

Within the nonlinear realization scenario, there are two operators which
contribute to anomalous couplings at lowest order\cite{DawVal}. They
correspond to ${\cal O}_B$ and ${\cal O}_W$ in Eq.~(\ref{ops}). In both,
the linear and non-linear realization scenarios, non-standard three
vector boson couplings are of ${\cal O}(m_W^2/\Lambda^2)$. If the
energy scale of the new physics is $\sim
1$~TeV, anomalous gauge boson couplings are expected to be no larger than
${\cal O}(10^{-2})$. Current high precision experiments still allow
anomalous couplings of ${\cal O}(1)$\cite{DPF}.

Tree level unitarity uniquely restricts the $WWV$ and $Z\gamma V$ couplings
to their
SM gauge theory values at asymptotically high energies\cite{CORNWALL}.
This implies
that either any deviation of $g_1^V$, $\kappa_V$, $\lambda_V$ and $h^V_i$,
$i=3,4$ has to
be described by a form factor, or that the effective Lagrangian
describing the anomalous vector boson self-interactions breaks down at
very high energies, $\sqrt{\hat s}$. The functional behaviour
of the form factors depends on the details of the underlying new
physics. Effective Lagrangian techniques are of little help here because
the low energy expansion which leads to the effective Lagrangian exactly
breaks down where the form factor effects become important. Therefore,
ad hoc assumptions have to be made. Here, we assume a behaviour similar
to the nucleon form factor
\begin{equation}\label{FF}
\Delta\kappa_V(\hat s) = { \Delta\kappa_V^0 \over
(1+\hat s/\Lambda_{FF}^2)^n }\; ,
\label{formf}
\end{equation}
and similarly for the other couplings. $\Lambda_{FF}$ in
Eq.~(\ref{formf}) is the form
factor scale which is a function of the scale of new physics, $\Lambda$.
We shall assume that $n=2$ for $WWV$ couplings, and $n=3$ ($n=4$) for
$h_3^V$ ($h_4^V$).
Since anomalous couplings are probed over a large $\hat s$ range in
di-boson production at hadron colliders, it is mandatory to take form
factor effects into account in these processes in order to avoid
unphysically large cross sections at high energies.

Some of the features of anomalous couplings, namely form factors and the
necessity to consider the full $S$-matrix elements can nicely be illustrated
by some very non-anomalous physics, namely fermion loop corrections within
the SM. At the same time the problem of how to
implement finite $W$ width effects while maintaining gauge invariance when
dealing with processes involving vector boson self-interactions can be
addressed.

Let us consider $W\gamma\to\ell\nu\gamma$ production at hadron
colliders. Replacing the $W$ propagator factors $1/(q^2-m_W^2)$ by a naive
Breit-Wigner form, $1/(q^2-m_W^2+im_W\Gamma_W)$, where $\Gamma_W$ is the
$W$ width, will disturb the gauge
cancellations between the individual Feynman graphs and thus lead to an
amplitude which is not electromagnetically gauge invariant. Finite
width effects are properly included by resumming the imaginary part of
the $W$ vacuum polarization. In the
unitary gauge and for $q^2>0$ the $W$ propagator is thus given by
\begin{equation}
D_W^{\mu\nu}(q)={-i \over q^2-m_W^2 + i q^2 \gamma_W}
\left( g^{\mu\nu}-{q^\mu q^\nu \over m_W^2}
(1+ i \gamma_W )   \right)\; ,  \label{Wprop}
\end{equation}
with $\gamma_W = \Gamma_W/m_W$. Note that
the $W$ propagator has received a $q^2$ dependent effective width which
actually would vanish in the space-like region. A gauge
invariant expression for $q\bar q'\to\ell\nu\gamma$ is obtained by
attaching the final state photon
in all possible ways to all charged particle propagators in the Feynman
graphs, including the charged fermions inside the $W$ vacuum
polarization loops\cite{BZ1}. As a result, for the process considered
here, the lowest order $WW\gamma$ vertex function
$\Gamma_0^{\alpha\beta\mu}$ is replaced by
\begin{equation}\label{WWgvertex}
\Gamma^{\alpha\beta\mu} =
\Gamma_0^{\alpha\beta\mu}\left(1+i\gamma_W \right)\; .
\end{equation}
The modification of the lowest order $WW\gamma$ vertex in
Eq.~(\ref{WWgvertex})
looks like the introduction of anomalous couplings $g_1^\gamma = \kappa_\gamma
= 1+i\gamma_W$ and one may thus worry that the full amplitude will violate
unitarity at large center of mass energies $\sqrt{\hat s}$. While indeed the
vertex is modified, this modification is compensated by the effective
$\hat s$-dependent width in the propagator. This interplay of
propagator and vertex corrections illustrates the remarks made above.
The leading one-loop contributions, namely the imaginary
parts of $WW\gamma$ vertex and inverse $W$ propagator, lead to a change of
the $S$-matrix element for $W\gamma$ production which can be
parameterized in terms of form factors by
\begin{equation}
g_1^\gamma(\hat s)=\kappa_\gamma(\hat s) = 1\; -\;
{i\Gamma_W m_W \over \hat s - m_W^2 + i\gamma_W \hat s}
\; , \label{ffexplicit}
\end{equation}
and the form factor scale is set by the $W$ mass.

\section{Tests at Hadron Colliders}
The signals of anomalous gauge boson couplings in di-boson production at
hadron colliders ($q \bar q \to W^+W^-,\; W\gamma,\;
Z\gamma,\; WZ$) can be understood from the high energy behaviour of the
anomalous contributions to the helicity amplitudes. Terms proportional
to non-standard couplings increase with energy like
$(\sqrt{\hat s}/m_W)^m$, where $m$ depends on the coupling and the
process considered\cite{DPF}, and $\hat s$ is the di-boson invariant
mass squared. For large values of $\sqrt{\hat s}$, the non-standard
contributions to the helicity
amplitudes would dominate, and would suffice to explain differential
distributions
of the photon and the $W/Z$ decay products. Due to the fact that
anomalous couplings only contribute via $s$-channel $W$, $Z$ or photon
exchange, the transverse momentum ($p_T$) distribution of the
vector boson should be particularly sensitive to non-standard $WWV$ and
$Z\gamma V$ couplings. This is demonstrated in Fig.~\ref{FIG:ONE},
where we show the $Z$ boson $p_T$ distribution in
$p\bar p\to W^+Z\to\ell_1^+\nu_1\ell_2^+\ell_2^-$, $\ell_{1,2}=e,\,
\mu$, at the Tevatron for the SM and various anomalous $WWZ$ couplings.
\setlength{\unitlength}{0.7mm}
\begin{figure}[t]
\begin{picture}(100,110)(0,1)
\centerline{\mbox{\epsfxsize5.8cm\epsffile{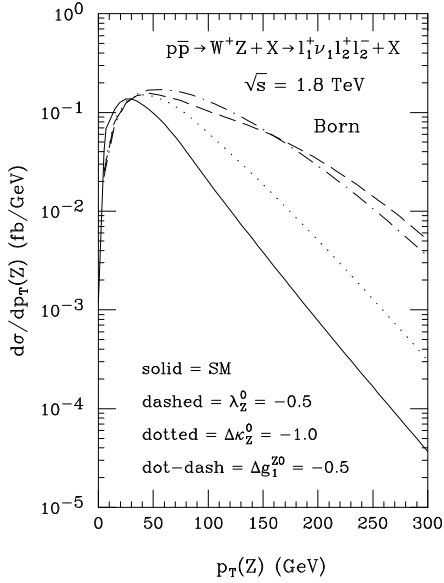}}}
\end{picture}
\caption{The differential cross section for the transverse momentum  of
the $Z$ boson in $p\bar p\to W^+Z$ at the Tevatron in the SM case
(solid line) and for various anomalous $WWV$ couplings. A form factor
scale of $\Lambda_{FF}=1$~TeV has been assumed.}
\label{FIG:ONE}
\end{figure}
Only one coupling is assumed to deviate from the SM at a time.
To simulate detector response, a $p_T(\ell),\, p\llap/_T>20$~GeV, and
a $|\eta(\ell)|<2.5$ cut have been imposed.

The current\cite{Dirk} limits on anomalous $WWV$ and $Z\gamma V$
couplings ($h_3^V=h_4^V=0$ in the SM at tree level) from Tevatron
experiments are summarized in Table~\ref{TAB:ONE}.
\begin{table}[hbt]
\begin{center}
\caption{95\% CL limits on anomalous $WWV$, $V=\gamma, \, Z$, and
$Z\gamma V$ couplings from current Tevatron experiments
($\ell_{1,2}=e,\,\mu$). Only one of the independent
couplings is allowed to deviate from the SM at a time. }
\label{TAB:ONE}
\vspace{0.2cm}
\begin{tabular}{c|c|c}
\hline\hline
experiment & channel & limit \\ \hline
$\!\!\!$CDF (prel.)$\!\!\!$ & $W\gamma\rightarrow\ell\nu\gamma$ &
$-1.8<\Delta\kappa_\gamma^0<2.0$ \\
67~pb$^{-1}$ & $\ell=e,\,\mu$ & $-0.7<\lambda_\gamma^0<0.6$ \\[1.mm] \hline
D\O\ & $W\gamma\rightarrow\ell\nu\gamma$ &
$-1.6<\Delta\kappa_\gamma^0<1.8$ \\
14~pb$^{-1}$ & $\ell=e,\,\mu$ & $-0.6<\lambda_\gamma^0<0.6$ \\ \hline
CDF & $\!\!\!WW,\, WZ\rightarrow \ell^\pm\nu jj\!\!\!$ &
$-1.0 <\Delta\kappa_V^0<1.0$ \\
20~pb$^{-1}$ & $\kappa_\gamma=\kappa_Z$,
$\lambda_\gamma=\lambda_Z$ & $-0.6<\lambda_V^0<0.7$ \\ \hline
D\O\ (prel.) & $\!\!\!WW,\, WZ\rightarrow e^\pm\nu jj\!\!\!$ &
$-0.9 <\Delta\kappa_V^0<1.1$ \\
14~pb$^{-1}$ & $\kappa_\gamma=\kappa_Z$,
$\lambda_\gamma=\lambda_Z$ & $-0.7<\lambda_V^0<0.7$ \\ \hline
D\O\ & $\!\!\!WW\rightarrow \ell_1\nu_1\ell_2\nu_2\!\!\!$ &
$-2.6 < \Delta\kappa_V^0 < 2.8$ \\
14~pb$^{-1}$ & $\kappa_\gamma=\kappa_Z$,
$\lambda_\gamma=\lambda_Z$ & $-2.2 < \lambda_V^0 < 2.2$ \\ \hline
$\!\!\!$CDF (prel.)$\!\!\!$ & $Z\gamma\rightarrow\ell^+\ell^-\gamma$ &
$-1.6<h^V_{30}<1.6$ \\
67~pb$^{-1}$ & $\Lambda_{FF}=0.5$~TeV & $-0.
4<h^V_{40}<0.4$ \\ \hline
D\O\ & $Z\gamma\rightarrow\ell^+\ell^-\gamma$ &
$-1.9<h^V_{30}<1.8$ \\
14~pb$^{-1}$ & $\Lambda_{FF}=0.5$~TeV & $-0.5<h^V_{40}<0.5$
\\ \hline\hline
\end{tabular}
\end{center}
\end{table}

For the Main Injector Era, integrated luminosities of the order of
1~fb$^{-1}$ are envisioned\cite{Tuts}, and through further upgrades of
the Tevatron
accelerator complex, an additional factor~10 in luminosity may be gained
(TeV33)\cite{jac}. The substantial increase in integrated luminosity
will make it
possible to test the $WWV$ and $Z\gamma V$ vertices with much greater
precision than in current experiments. In Fig.~\ref{FIG:TWO} we compare
the limits on $WWV$ couplings expected from $e^+e^-\to
W^+W^-\to\ell\nu jj$ and
the various di-boson production processes in hadronic collisions
in the HISZ scenario [see Eqs.~(\ref{EQ:HISZ1})
--~(\ref{EQ:HISZ3})] for the envisioned energies and integrated
luminosities of the Tevatron and LEP~II\cite{BHO1}.
Similar bounds are obtained if different relations between the anomalous
couplings are assumed.
\setlength{\unitlength}{0.7mm}
\begin{figure}[thb]
\begin{picture}(100,110)(0,1)
\centerline{\mbox{\epsfxsize7.1cm\epsffile{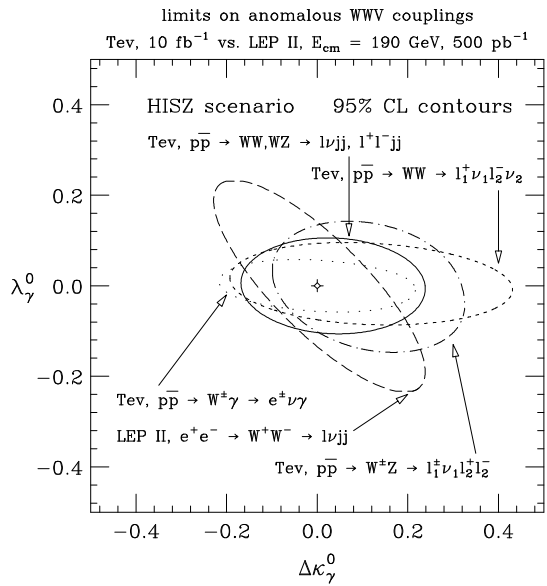}}}
\end{picture}
\caption{Comparison of the expected sensitivities on anomalous $WWV$
couplings in the HISZ scenario [see Eqs.~(\protect{\ref{EQ:HISZ1}})
--~(\protect{\ref{EQ:HISZ3}})] from $e^+e^-\to
W^+W^-\to\ell^\pm\nu jj$ at LEP~II and various processes at the Tevatron.}
\label{FIG:TWO}
\end{figure}
The bounds expected from Tevatron and LEP~II data for
$\Delta\kappa_\gamma$ are quite similar ($\approx 0.2$), whereas the
Tevatron enjoys a clear advantage in constraining
$\lambda_\gamma$ ($|\lambda_\gamma^0|<0.06$), if correlations between
the two couplings are taken
into account. It should be noted, however, that the strategies to extract
information on vector boson self-interactions at the two machines are
very different. At the
Tevatron one exploits the strong increase of the anomalous
contributions to the helicity amplitudes with energy to derive limits.
At LEP~II, on the other hand, information is extracted from the angular
distributions of the final state fermions. Data from the Tevatron and
LEP~II thus yield complementary information on the nature of the $WWV$
couplings.

Because of the much higher energies accessible at the Tevatron and the
steep increase of the anomalous contributions to the helicity
amplitudes with energy, Tevatron experiments will be able to place
significantly better bounds on the $Z\gamma V$ couplings ($|h^V_{30}|<0.
024$, $|h^V_{40}|<0.0013$ for $\Lambda_{FF}=1.5$~TeV and $\int\!{\cal
L}dt=10$~fb$^{-1}$) than LEP~II ($|h^V_{30,40}|<0.5$)\cite{DPF,TEV33}. The
Tevatron limits,
however, do depend non-negligibly on the form factor scale assumed.

Since terms proportional to the non-standard $WWV$ and $Z\gamma V$
couplings in the
di-boson production amplitudes grow quickly with energy, one
expects\cite{BZ2} that experiments at the LHC are able to improve
significantly the bounds which can be obtained at the Tevatron. The
effect of QCD corrections, which are very large in di-boson production
processes at the LHC\cite{BHO1,BHO2}, and the $t\bar t$, $t\bar t\gamma$
and $t\bar tZ$ backgrounds can be reduced by imposing a jet
veto\cite{DPF}. Table~\ref{TAB:TWO} compares
the sensitivities which can be achieved in di-boson
production at the LHC with 100~fb$^{-1}$ for $\Lambda_{FF}=3$~TeV. If
the integrated luminosity
is reduced by a factor~10, the bounds listed in Table~\ref{TAB:TWO} are
weakened by
about a factor~2. The limits depend non-negligibly on the form factor
scale assumed; for $\Lambda_{FF}>1$~TeV, $\Delta\kappa_V$ and
$\lambda_V$ can be probed to better than 0.1 and 0.01, respectively.
\begin{table}[t]
\begin{center}
\caption{Expected 95\% CL limits on anomalous $WWV$, $V=\gamma, \, Z$,
and $Z\gamma V$ couplings from experiments at the LHC with 100~fb$^{-1}$
and $\Lambda_{FF}=3$~TeV ($\ell_{1,2}=e,\, \mu$). Only one of the
independent couplings is assumed to deviate from the SM at a time. A
jet veto and standard lepton identification cuts are imposed to reduce
backgrounds.}
\label{TAB:TWO}
\vspace{0.2cm}
\begin{tabular}{c|c}
\hline\hline
channel & limit \\ \hline
$W\gamma\rightarrow e\nu\gamma$ &
$\!\!\!\!\!\!\!\!-0.080<\Delta\kappa_\gamma^0<0.080$
\\
& $\!\!\!\!\!\!\!\!-0.0057<\lambda_\gamma^0<0.0057$
\\ \hline
$WZ\to\ell_1\nu_1\ell_2^+\ell_2^-$ & $\!\!\!\!\!\!\!\!-0.0060
<\Delta\kappa_\gamma^0<0.0097$ \\
HISZ scenario &
$\!\!\!\!\!\!\!\!-0.0053<\lambda_\gamma^0<0.0067$
\\ \hline
$WZ\to\ell_1\nu_1\ell_2^+\ell_2^-$ & $\!\!\!\!\!\!\!\!-0.064
<\Delta\kappa_Z^0<0.107$ \\
$\Delta g_1^Z=0$ &
$\!\!\!\!\!\!\!\!-0.0076<\lambda_Z^0<0.0075$
\\ \hline
$WW\to\ell_1\nu_1\ell_2\bar\nu_2$ & $\!\!\!\!\!\!\!\!-0.
025 < \Delta\kappa_\gamma^0 < 0.047$ \\
HISZ scenario & $\!\!\!\!\!\!\!\!-0.0079<
\lambda_\gamma^0<0.0078$
\\ \hline
$WW\to\ell_1\nu_1\ell_2\bar\nu_2$ & $\!\!\!\!\!\!\!\!-0.
018 < \Delta\kappa_Z^0 < 0.027$ \\
$\kappa_\gamma=g_1^Z=1$, $\lambda_\gamma=0$ &
$\!\!\!\!\!\!\!\!-0.0111<\lambda_Z^0<0.0084$
\\ \hline
$Z\gamma\rightarrow e^+e^-\gamma$ &
 $\!\!-0.0013<h^V_{30}<0.0013$ \\ &
$\!\!-6.8\cdot 10^{-6}<h^V_{40}<6.8\cdot 10^{-6}$ \\ \hline\hline
\end{tabular}
\end{center}
\end{table}
The sensitivities which are expected for $Z\gamma V$ couplings at the
LHC depend very strongly on $\Lambda_{FF}$. For
$\Lambda_{FF}=3$~TeV, $h^V_{30}$ ($h^V_{40}$) can be probed at the
${\cal O}(10^{-3})$ (${\cal O}(10^{-5})$) level. Reducing the form factor
scale by a factor~2, weakens the limits by a factor five to ten\cite{BB}.

\section{Conclusions}
We have discussed the theoretical aspects of $WWV$ and $Z\gamma V$
couplings and their measurement at the Tevatron and LHC. These
couplings are defined through a phenomenological effective Lagrangian,
analogously to the
general vector and axial vector couplings, $g_V$ and $g_A$, for the
coupling of gauge bosons to fermions. The major goal of these
measurements will be the confirmation of the SM
predictions. If the energy scale of the
new physics responsible for the non-standard gauge boson couplings is $\sim
1$~TeV, these couplings are expected to be no larger than
${\cal O}(10^{-2})$.

Present data from di-boson production at the
Tevatron yield bounds typically in the range of 0.5~--~2.0.
Within the next 10~years, experiments conducted at the Tevatron and at
LEP~II are expected to confirm the SM $WWV$ ($Z\gamma V$) couplings at the
10\% (1\%) level. At the LHC one expects to probe anomalous $WWV$
couplings with a precision of ${\cal O}(10^{-1}-10^{-3})$. The limits
on the $Z\gamma V$ couplings are very sensitive to the value of
$\Lambda_{FF}$. For
$\Lambda_{FF}\ge 3$~TeV, the bounds which can be achieved are of ${\cal
O}(10^{-3})$ for $h_3^V$, and of ${\cal O}(10^{-5})$ for $h_4^V$.

\setcounter{secnumdepth}{0} 


\end{document}